\documentclass[prd,preprint,showpacs,showkeys,amsmath,amssymb,superscriptaddress]{revtex4}%

\usepackage[dvips]{graphics}
\usepackage{array,amsmath,amssymb}
\usepackage{amsmath}
\usepackage{amssymb}

\begin{document}


\title{Properties of Color-Coulomb String Tension}

\author{Y.~Nakagawa} 
\affiliation{Research Center for Nuclear Physics, 
 Osaka University, Ibaraki, Osaka 567-0047, Japan}

\author{A.~Nakamura }
\affiliation{Research Institute for Information Science and Education, Hiroshima University, Higashi-Hiroshima 739-8521, Japan}

\author{T.~Saito}
\affiliation{Research Center for Nuclear Physics, 
 Osaka University, Ibaraki, Osaka 567-0047, Japan}

\author{H.~Toki}
\affiliation{Research Center for Nuclear Physics, 
 Osaka University, Ibaraki, Osaka 567-0047, Japan}

\author{D.~Zwanziger}
\affiliation{Physics Department, New York University,
 New York, NY~10003, USA}

\begin{abstract}
We study the properties of the color-Coulomb string tension 
obtained from the instantaneous part of gluon propagators in Coulomb gauge  
using quenched $SU(3)$ lattice simulation.
 In the confinement phase, 
the dependence of the color-Coulomb string tension 
on the QCD coupling constant is smaller than 
that of the Wilson loop string tension.
On the other hand, 
in the deconfinement phase, 
the color-Coulomb string tension 
 does not vanish even for $T/T_c = 1 \sim 5$, 
 the temperature dependence of which is comparable with the magnetic scaling, 
 dominating the high temperature QCD. 
Thus, 
the color-Coulomb string tension is not an order parameter of 
 QGP phase transition. 
\end{abstract}

\pacs{12.38.Aw, 12.38.Gc, 11.15.Ha}
\keywords{lattice QCD, color confinement, Coulomb gauge, quark-gluon plasma}

\maketitle

\maketitle

\section{Introduction}

Understanding color confinement in quantum chromodynamics (QCD)
 is one of the most challenging problems in quantum field theory, and 
also provides essential knowledge for low temperature hadron
physics. There are many approaches to understand 
color confinement dynamics:
 dual superconductor scenario, center vortex model, 
 the infrared behavior of gluon propagators, etc.
 have been widely studied and a large amount of information 
 on color confinement
 has been accumulated. See reviews in Refs. \cite{TR,NR,GR}.
 In those scenarios, topological 
objects and gauge-dependent quantities,
 bringing out the properties of QCD vacuum,
 may play an important role. 
A key issue is the choice of gauge in which 
the confinement scenario is realized.
 
Recently, there has been considerable interest
 in the Coulomb gauge color confinement scenario.
 This scenario was originally discussed by Gribov \cite{Gribov}, and 
 in recent years, Zwanziger has advocated the importance of a color-Coulomb
potential in Coulomb gauge for color confinement \cite{CCG}.
He and his collaborators showed that, in Coulomb gauge,
the time-time component of gluon propagators, $g^2 D_{00}$,
including an instantaneous color-Coulomb potential plus
a noninstantaneous vacuum polarization,
is invariant under renormalization \cite{CCG,PRC,renorm}.
It has been found by perturbative analysis \cite{renorm} that 
the instantaneous part in Coulomb gauge QCD causes antiscreening, 
while the vacuum polarization part causes screening. Hence, one expects that
the instantaneous color-Coulomb potential represents a linearly rising 
behavior for large quark separations. 
Moreover, Zwanziger pointed out that there is
an inequality  \cite{Zwan},
$ V_{phys}(R) \le V_{coul}(R), \label{ZwanzigerIneq} $
where $V_{phys}(R)$ means a physical heavy-quark-antiquark potential
and $V_{coul}(R)$ the Coulomb heavy-quark potential
 corresponding to the instantaneous part of $D_{00}$.
This inequality indicates that
if the physical heavy-quark potential is confining,
then the Coulomb heavy-quark potential is also confining.
See Ref. \cite{DZ-ptps98} for a review.

In order to verify the Coulomb gauge color confinement scenario, 
one needs a nonperturbative technique to describe
 the low-energy color confinement.
 Therefore, nonperturbative verifications 
have been tried in lattice gauge simulation.
In the $SU(2)$ lattice numerical simulation carried out
by Cucchieri and Zwanziger \cite{minimal},
it was found that $g^2 D_{00}(\vec{k})$
is strongly enhanced at $\vec{k}=0$.
In $SU(2)$ and $SU(3)$ lattice simulations 
\cite{Greensite,Greensite2,PPLC3}, furthermore,
it was reported that the Coulomb heavy-quark potential
grows linearly at large quark separations in the confinement phase.

On the lattice,   
it is essential to study the magnitude and 
the scaling for the string tension, which is a characteristic quantity
for confinement physics.
Numerical lattice calculations \cite{Greensite,Greensite2,PPLC3}
 indicate that the color-Coulomb string tension has $2-3$ times 
larger value in comparison with the case of a gauge invariant 
Wilson loop, as expected by 
$ V_{phys}(R) \le V_{coul}(R)\label{ZwanzigerIneq}$.
In addition, 
the $SU(2)$ lattice numerical data in Ref. \cite{Greensite2} 
 show the possibility that an asymptotic scaling violation
 for the color-Coulomb string tension may be less 
 than the usual Wilson loop string tension.  
Accordingly, the dependence of the 
color-Coulomb string tension on a gauge coupling or a lattice cutoff
ought to be extensively investigated in $SU(3)$ lattice gauge theory.

The lattice simulations mentioned above have
 shown the linearity of the instantaneous color-Coulomb potential
 at large distances in the confinement regions.
At the same time,
 the lattice calculations at finite temperature 
in the deconfinement phase
 indicate that the Coulomb string tension remains
after the quark-gluon plasma (QGP) phase transition \cite{Greensite2,PPLC3}.
One possible explanation is that
the color-Coulomb potential, 
 determined by the spatial-like and 
time (temperature) independent Faddeev-Popov operator,  
 is not sensitive to the system temperature. 
 In addition, 
 we note that  
 the potential obtained from a spatial Wilson loop
 above $T_c$ behaves as a linearly rising function
\cite{SWL1,SWL2,SWL3,SWL4,SWL5,SWL6,SWL7,SWL8}.
Both the color-Coulomb and the spatial Wilson potentials 
have a common feature that
they are defined by spatial variables.
However, the higher temperature lattice simulations than 
the previous calculations \cite{Greensite2,PPLC3}
are indispensable.
Nevertheless, 
the noninstantaneous retarded part with the vacuum polarization 
still gives a color-screened potential \cite{PPLC3,CDP1,CDP2}.    

In Coulomb gauge QCD there are no unphysical degrees of freedom 
for gauge fields; namely, Coulomb gauge is a physical gauge.
In contrast, Lorentz covariant gauges generate a negative spectral function
 due to the indefinite metric of Fock space. 
This is very convenient on discussing a physical hadron, 
and a lot of attempts 
 have been made to construct models 
 based on the Coulomb gauge Hamiltonian to describe 
 color confinement \cite{M1,M2,M3,M4,M5} and hadrons \cite{M6,M7}.

In this paper, we will perform more extensive lattice QCD studies on 
the Coulomb gauge confinement scenario
 comparing with the previous calculation \cite{PPLC3}.
In the confinement phase, we investigate the scaling behavior 
of the color-Coulomb string tension by varying a lattice cutoff or
a coupling constant $\beta=6/g^2$.
In the deconfinement phase, we discuss the relation between 
the thermal color-Coulomb string tensions, which are calculated 
at high temperatures, $T/T_c = 1.5 \sim 5.0$, and
 the magnetic scaling that is believed to dominate the high temperature QCD. 
In Sec. II we briefly review the partition function in Coulomb gauge and 
describe the instantaneous color-Coulomb potential and 
the noninstantaneous vacuum polarization part.
In Sec. III, we give the definition of
 the partial-length Polyakov line correlator\cite{Greensite,Greensite2} 
 to evaluate the instantaneous part.
Section IV is devoted to show the numerical results.
Section V gives conclusions.

\section{Instantaneous color-Coulomb potential}

The construction of 
the partition function in Coulomb gauge 
through the Faddeev-Popov technique and the 
derivation of the instantaneous color-Coulomb potential 
were done in Ref. \cite{PRC}. 
The Hamiltonian of QCD in Coulomb gauge can be written as
\begin{equation}
\begin{array}{ccc}
H = \displaystyle
    \frac{1}{2} \int d^3 x ( E_i^{tr2}(\vec{x})+B_i^2(\vec{x}) ) +
    \displaystyle
    \frac{1}{2} \int d^3 x d^3 y
    (\rho(\vec{x}){\cal V}(\vec{x},\vec{y}) \rho(\vec{y}) ),
\label{part}
\end{array}
\end{equation}
where $E_i^{tr}$, $B_i$ and $\rho$ are
  the transverse electric field, the magnetic field and
 the color charge density, respectively.
The function $\cal V$ in the second term 
 is made by the Faddeev-Popov (FP) operator in the spatial direction, 
 $M=-\vec{D} \vec{\partial}=
 -(\vec{\partial}^2 + g\vec{A} \times \vec{\partial})$, 
\begin{equation}
{\cal V}(\vec{x},\vec{y})= \int d^3 z 
\left[ \frac{1}{M(\vec{x},\vec{z})}
 ( -\vec{\partial}^2_{(\vec{z})} )
   \frac{1}{M(\vec{z},\vec{y})} \right].
\end{equation} 
 
From the partition function with Hamiltonian Eq. (\ref{part}),   
 one can evaluate the time-time gluon propagator composed 
of the following two parts:
\begin{equation}
 g^2\langle A_0(x) A_0(y) \rangle = 
 g^2 D_{00}(x-y) = V (x-y) + P (x-y), 
\label{d00}
\end{equation}
where 
\begin{equation}
V(x-y) = g^2 \langle {\cal V}(\vec{x},\vec{y}) \rangle \delta(x_4-y_4). 
\label{Vi}
\end{equation}
The equation (\ref{Vi})
 corresponds to the instantaneous color-Coulomb potential
at equal time and causes antiscreening; 
namely, it is the most important quantity in Coulomb gauge confinement
scenario, and is constructed by the spatial FP matrix.
Therefore, if the potential $V$ is a linearly rising potential for large 
quark separations, then color confinement is attributed to 
an enhancement of the low-lying mode of FP eigenvalues \cite{Gribov,CCG}.
Note that Eq. (\ref{Vi}) in the case of 
quantum electrodynamics (QED) as a non-confining theory 
is identified as a Coulomb propagator $\langle -1/\partial_i^2 \rangle$
or a Coulomb potential $ 1/r $.
Simultaneously, 
the quantity $P$ in Eq. (\ref{d00}) is a vacuum polarization term,
\begin{equation}
P(x-y) = -g^2 
\langle 
\int {\cal V}(\vec{x},\vec{z})\rho(\vec{z},x_4) d^3 z 
\int {\cal V}(\vec{y},\vec{z}{\,'})\rho(\vec{z}{\,'},y_4) d^3 z' 
 \rangle,
\end{equation}
which causes color-screening effect
owing to the minus sign of this equation, 
 and produces the reduction of a color-confining force and 
a quark-pair creation from vacuum when dynamical quarks exist.
Moreover, this perturbative argument is also satisfied at 
one-loop order \cite{CCG}. 

\section{Partial-length Polyakov line }

In this section, we give the definition of 
a static heavy quark-antiquark potential 
 in the color-singlet channel as a function of distance, $R$, and  
summarize how to fix the gauge on the lattice.

We introduce a partial-length Polyakov line (PPL) defined as
\cite{Greensite,Greensite2}
\begin{equation}
L(\vec{x},n_t) = \displaystyle\prod_{n_s=1}^{n_t} U_0(\vec{x},n_s),
\quad n_t=1, 2, \cdots, L_t.
\end{equation}
Here $U_0(\vec{x},t)= \exp(iagA_0(\vec{x},t))$
is an $SU(3)$ link variable in the temporal direction and 
$a$, $g$, $A_0(\vec{x},t)$ and $L_t$ represent the lattice cutoff,
the gauge coupling, the time component of gauge potential and
the temporal-lattice size. 
A PPL correlator in color-singlet channel is given by
\begin{equation}
G(R,n_t) = \frac{1}{3}\left< Tr[L(R,n_t)L^{\dagger}(0,n_t)] \right>,
\label{pots}
\end{equation}
where $R$ stands for $\arrowvert \vec{x} \arrowvert $.
From Eq. (\ref{pots}) one evaluates the color-singlet potential
on the lattice, 
\begin{equation}
V(R,n_t) = \log \left[
\frac{G(R,n_t)}{G(R,n_t+1)} \right]\label{pot1}. 
\end{equation}
In the case of $n_t=0$,
we define
\begin{equation}
V(R,0) = - \log [ G(R,1) ]\label{pot2}.
\end{equation}

Greensite et al. argued that
this function $V(R,0)$ in Coulomb gauge corresponds 
to an instantaneous color-Coulomb potential $V_{coul}(R)$ 
 \cite{Greensite,Greensite2}. 
The potential $V(R,n_t)$ in the limit $n_t \rightarrow \infty$
 is expected to correspond to a physical potential,
$V_{phys}(R)$, usually calculated from 
the Wilson loops in the same limit.
Both potentials are known to satisfy 
Zwanziger's inequality, $V_{phys}(R) \le V_{coul}(R)$ \cite{Zwan}.

Since the color-decomposed potential defined by the PPL correlator  
such as Eq. (\ref{pots}),
do not have a gauge invariant form, 
we must fix the gauge. One can realize the Coulomb gauge on the lattice
to maximize the measurement 
\begin{equation}
 \sum_{\vec{x}} \sum_{i=1}^3
\mbox{ReTr} U_i^{\dagger}(\vec{x},t), 
\end{equation}
by repeating the gauge rotations:
\begin{equation}
U_i(\vec{x},t) \rightarrow U_i^{\omega}(\vec{x},t)
= \omega^{\dagger}(\vec{x},t)U_i(\vec{x},t) 
\omega(\vec{x}+\hat{i},t),
\end{equation}
where $\omega$ $\in SU(3)$
\footnote{Here, we adopt $\omega=e^{i\alpha\partial_i A_i}$ as 
the gauge rotation matrix and suitably chose the parameter $\alpha$, 
which depends mainly on the lattice size.}
 is a gauge rotation matrix
 and $U_i(\vec{x},t)$ are link variables for the spatial direction.
Thus, each lattice configuration thermalized 
after the Monte Carlo quantization can be gauge fixed iteratively
\cite{Mandula}.

\section{Results and discussions}

We carried out $SU(3)$ lattice gauge simulations
in the quenched approximation to calculate 
the instantaneous color-Coulomb $q\bar{q}$ potential  
 in the confinement and deconfinement phases.
The lattice gauge configurations were generated by
 the standard heat-bath Monte Carlo technique
 with a simple plaquette Wilson gauge action.

\subsection{Linearity of instantaneous color-Coulomb potential}

\begin{figure}[htbp]
\begin{center}
\resizebox{10cm}{!}{\includegraphics{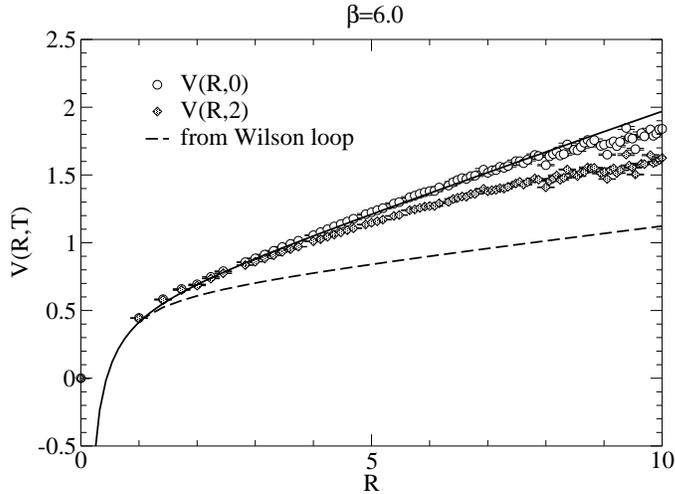}}
\caption{An example of the R-dependence of
 the instantaneous color-Coulomb potential
 (in dimensionless lattice units)
 at $\beta=6.0$ ($a\sim 0.1$ fm)
 on a $18^3 \times 32$ lattice in the confinement phase.
This result was obtained in the previous work \cite{PPLC3}.
The solid and dashed curves stand for the fitted result for the 
potential $V(R,0)$ and the Wilson loop potential reported 
in Ref. \cite{Bali}, respectively.
} 
\label{CPotentialatzero}
\end{center}
\end{figure}

An example of the variation of the instantaneous color-Coulomb 
 potential $V(R,0)$ with distances is shown in Fig. \ref{CPotentialatzero},
 which demonstrates that 
the potential $V(R,0)$ behaves as a linearly rising function 
with increasing distance $R$ and
 can be described in terms of the Coulomb term plus 
linear term with a nonzero string tension, 
\begin{equation}
V(R,0) = c_0 + KR + e/R \label{fitting},
\end{equation}
where $e$ is fixed to $-\pi/12$ for a two-parameter fit, and
 $K=\sigma_c a^2$ is
 the color-Coulomb string tension.
Thus we find that
the instantaneous potential $V(R,0)$ is a confining potential.
In contrast, 
the vacuum polarization (retarded) part causes color screening, which 
weakens the confining force as reported in Refs. \cite{Greensite,PPLC3}.
Consequently, the color-Coulomb potential 
in the limit $n_t \rightarrow \infty$ is 
expected to approach the Wilson loop potential.
The slope of the potential $V(R,n_t)$ with finite $n_t$ decreases
as displayed in Fig. \ref{CPotentialatzero}.
Note that
the numerical result in Fig. \ref{CPotentialatzero} 
was obtained in the previous work \cite{PPLC3},
and in the present study, 
we will not enter into details 
on the vacuum polarization part any further.

\subsection{Color-Coulomb string tension}

The string tension is a characteristic quantity in discussing confinement
physics and thus one should investigate 
the scaling behavior of the color-Coulomb string tension 
 $\sqrt{\sigma_c}=\sqrt{K}a^{-1}$, 
obtained by the lattice simulation with a finite cutoff.
Here we can introduce a two-loop asymptotic scaling of QCD with
 the mass parameter $\Lambda$ and lattice cutoff $a$ as 
\begin{equation}
a \Lambda = \exp \left( -\frac{1}{2b_0g^2} \displaystyle \right) 
(b_0 g^2)^{-\frac{b_1}{2b_0^2}} = f(g), 
\end{equation}
where $b_0$ and $b_1$ are universal first two coefficients of 
$\beta$ function. Since the quantity $\sqrt{\sigma_c}$ is expected 
to be proportional to 
the scale of QCD $\Lambda$ in asymptotic regions,   
it makes sense to consider the following relation: 
\begin{equation}
\frac{\sqrt{\sigma_c}}{\Lambda} = 
\frac{\sqrt{K}}{f(g)}, 
\end{equation}
which would be reduced to a constant in the continuum (weak coupling) limit.

\begin{figure}[htbp]
\begin{center}
\resizebox{10cm}{!}{\includegraphics{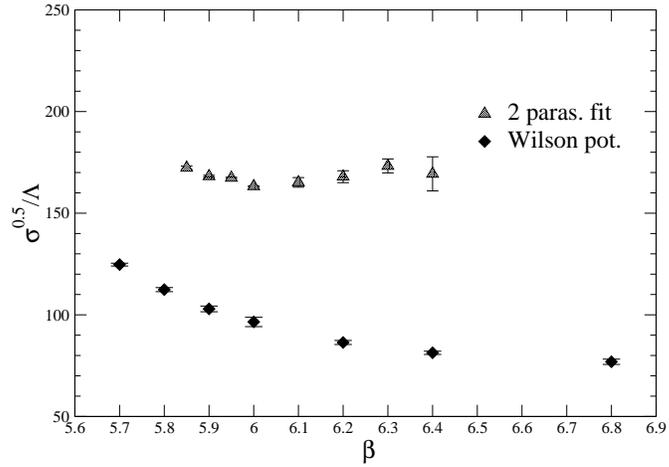}}
\caption{
The dependence of the color-Coulomb string tensions on $\beta=6/g^2$.
 The triangle symbols with error bar
 stand for the color-Coulomb string tension
 in the deconfinement phase.
The Wilson loop string tensions, represented 
by the symbols of the diamond shape, are also plotted for comparison 
taken from Table III in Ref. \cite{Bali-asym}. } 
\label{Stringatzero}
\end{center}
\end{figure}

In the present study, 
we carried out calculations at $\beta=6.1-6.4$ on a $18^4$ lattice
and used 300 gauge configurations measured every 100 sweeps after 
sufficient thermalization.
The $\beta=6/g^2$ dependence of the color-Coulomb string tensions
 is plotted in Fig. \ref{Stringatzero}, 
in which we additionally employed the data at $\beta=5.85-6.00$
 reported in the previous calculation \cite{PPLC3}.
 For the two-parameter fitting by Eq. (\ref{fitting}),  
we employed the data over ${\cal R} \sim 0.2$ fm
 up to ${\cal R} \sim 0.5$ fm for $\beta=6.1 - 6.4$, which are 
 $R \sim 3 - 6$ in lattice units,
 restricted due to the periodic boundary condition.     
Although the results of the high $\beta$ regions have 
large errors in our calculations
\footnote{
We also attempted a 3-parameter fit to Eq. (\ref{fitting}), 
keeping $e$ as a fitting parameter. However we have not plotted
the result of the fit because of the large uncertainty, 
particularly at high $\beta$.},
 the variations of the color-Coulomb string tensions as $\beta$ varies
seem to be smaller 
 than the case of the Wilson loop string tension ($\sqrt{\sigma_w}$)
\cite{Bali-asym}, included for comparison.
The relative fluctuation (the ratio of the minimum and maximum)
of those data is within $\sim 6(2)$ \%. 
Such tendency was also observed in the $SU(2)$ lattice
 simulations \cite{Greensite2}.

Moreover, in the range of $\beta$ used here, 
the value of $\sqrt{\sigma_c}$ still remains
 approximately 2 times as large as that of $\sqrt{\sigma_w}$, and 
monotonically varies with $\beta$.
This tendency seems to be unchanged in the continuum limit. 
However, the larger lattice simulation at higher $\beta$ is required 
to realize the asymptotic scaling. 

\subsection{Behavior of instantaneous color-Coulomb potential at $T\neq0$}

\begin{figure}[htbp]
\begin{center}
\resizebox{10cm}{!}{\includegraphics{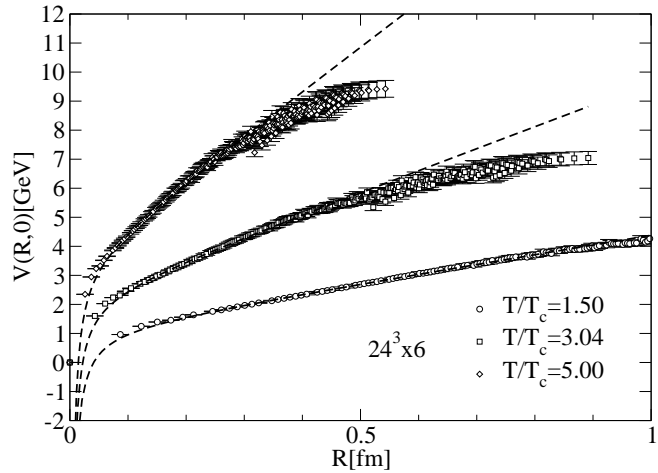}}
\caption{The dependence of 
the instantaneous color-Coulomb potential on 
the temperature in the deconfinement phase.
The dashed curves stand for the fitted results.} 
\label{CStringatnonzero}
\end{center}
\end{figure}

\begin{figure}[htbp]
\begin{center}
\resizebox{10cm}{!}{\includegraphics{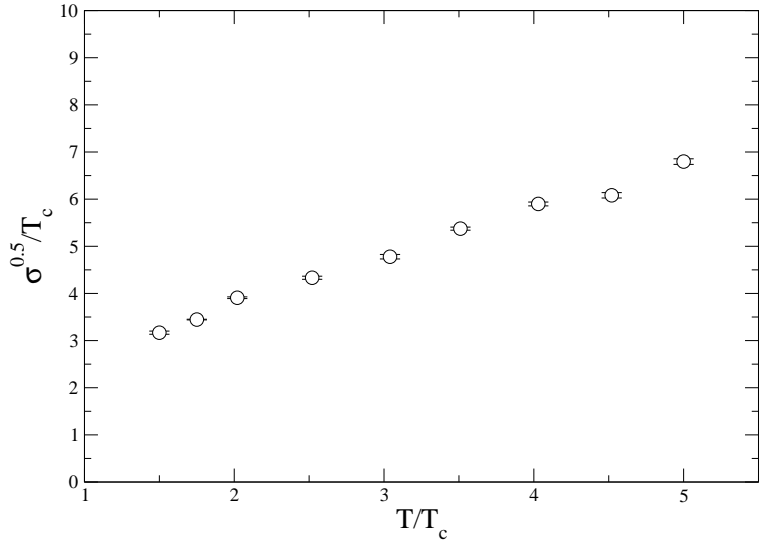}}
\caption{The temperature dependence of  
the color-Coulomb string tension in units of $T_c$
 in the deconfinement phase.
The color-Coulomb string tension is proportional to $T$. 
\label{StringvsT}
} 

\label{Beta-dep}
\end{center}
\end{figure}
In the confinement phase, as seen in the previous section, 
 the instantaneous color-Coulomb part gives a confining potential.
However, it is reported in Refs. \cite{Greensite2,PPLC3} 
that the linearity of $V(R,0)$ is not lost even after the QGP phase 
transition. Therefore, in the present work, 
we carried out lattice simulations at $T/T_c=1.5-5.0$
on the fixed lattice size $24^3 \times 6$.
 Here, the critical temperature of the QGP phase transition $T_c$ is 
approximately 256 MeV for $N_t=6$ \cite{Boyd}.
We fixed the lattice temperature  $T=1/N_t a$
to vary the lattice cutoff $a$ ($\beta$) \cite{qcdtaro}
 and 300 gauge configurations measured every 100 sweeps were used.
In Fig. \ref{CStringatnonzero}, we show the temperature dependence 
of the {\it thermal} color-Coulomb potential
 with a nonvanishing string tension.
The potentials $V(R,0)$ in the deconfinement phase still behave
 like a linear-confining potential, and furthermore
 the slope and magnitude of those potentials become larger.
It is found in Fig. \ref{StringvsT} that 
the main temperature dependence of the {\it thermal} color-Coulomb string
 tension is apparently given by the linear relation, 
\begin{equation}
\sqrt{\sigma_c} \sim T,
\end{equation}
 being directly proportional to the temperature.

\subsection{Spatial Wilson loop potential and magnetic scaling}

In order to interpret the confining phenomenon caused by 
the instantaneous color-Coulomb part in the deconfinement phase,
it may be instructive to review studies on the thermal behavior
of the {\it spatial} Wilson loop above $T_c$
 \cite{SWL1,SWL2,SWL3,SWL4,SWL5,SWL6,SWL7,SWL8}.
The spatial Wilson loop, $W(R,S)$, is constructed
 by the only spatial links (or spatial gluon fields),
 where $R$ and $S$ are spatial extents on a lattice. 
If the loop $W(R,S)$ follows area law as a function of $S$, then 
the spatial Wilson loop potential $V_s(R)$ would be given by 
\begin{equation}
V_s(R) = \lim_{S \rightarrow \infty} 
\ln \frac{W(R,S)}{W(R,S+1)}.
\end{equation}
The potential $V_s$ on the confined $T=0$ (symmetric hypercube) lattices 
can be identical with the usual Wilson loop potential.
However, in the deconfinement phase, it is known that the potential $V_s$ 
 is a linearly rising potential at large distances rather than 
 a color-screened potential. Consequently, 
the nonvanishing spatial string tension, $\sqrt{\sigma_s}$, exists even in 
the deconfinement phase.

In a series of studies on the thermal spatial Wilson loop, 
 the spatial string tension $\sqrt{\sigma_s}$ 
at high temperature 
has been discussed in terms of the magnetic scaling.  
According to perturbation theory of thermal QCD (TQCD) \cite{TQCD},
 the temporal gluon propagator yields the electric mass,
\begin{equation}
 m_e \sim g(T) T, \label{ele}
\end{equation}
 usually referred to as a color-Debye screening mass, 
 while the spatial gluon propagator also yields 
 the magnetic mass
\begin{equation}
m_m \sim g^2(T) T, \label{mag}
\end{equation}
 which must be introduced 
 as the cutoff factor to solve an infrared divergence that appears 
in TQCD perturbation.
 Thus, the magnetic scaling somewhat has a nonperturbative origin and
 is closely relevant to longer range physics than the electric scale.

The infrared sensitivity in TQCD is known to survive 
in the high temperature limit ($T\rightarrow \infty$) through 
the argument of 3-dimensional reduction \cite{3D1,3D2}.
This approach enables us to obtain the effective theory 
 that is defined by integrating out a nondynamical 
heavy mode in the high temperature limit; this theory 
proves that
the long-range properties of TQCD
 are dominated by the magnetic scaling.  

\subsection{Thermal color-Coulomb string tension}

To obtain the thermal color-Coulomb string tension $\sqrt{\sigma_c(T)}$
for the deconfining 
phase, we employed an ansatz of the Coulomb plus linear terms.
 The actual fitting analyses
 by the use of the same function as Eq. (\ref{fitting}) 
 give $\chi^2/ndf \lesssim 1 $ for the data of the fixed range of $R=3-7$
for $T/T_c=1.50-5.00$.
The thermal color-Coulomb string tensions 
do not vanish for those temperatures, the values of which  
 increase with temperature,
 and the rate of increase of the temperature is more rapid than 
those of the string tension $\sqrt{\sigma_c(T)}$.
This tendency is acceptable if the 
color-Coulomb string tension at finite temperature 
is regarded as a thermal quantity, such as  
 the electric and magnetic scaling described in Eqs. (\ref{ele}) and 
(\ref{mag}).

\begin{figure}[htbp]
\begin{center}
\resizebox{10cm}{!}{\includegraphics{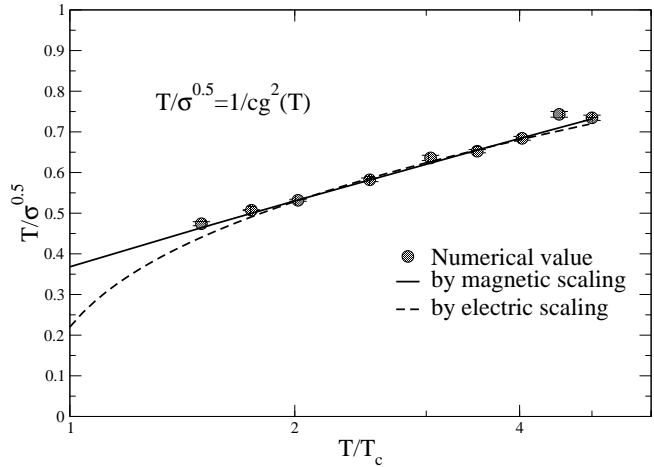}}
\caption{The dependence of the color-Coulomb string tension 
as a function of $T/T_c$.
The circle symbols are numerical data and 
the solid line represents the fitted result by 
the magnetic scaling, while the dashed curve (line) 
represents that by the electric scaling.
} 
\label{CStringatfinite}
\end{center}
\end{figure}

From Eq. (\ref{Vi}), it is clear that 
the instantaneous part made by the Faddeev-Popov 
matrix is independent of time and a spatial-like quantity
although the temporal and spatial gluon fields are correlated by 
the self-interaction in QCD.
This situation is very similar to the case of the spatial Wilson loop
at finite temperature as reviewed in the previous section.
Therefore, we shall describe these data by the magnetic scaling:
\begin{equation} 
\frac{T}{\sqrt{\sigma_c(T)}}=\frac{1}{c}\frac{1}{g^2(T)},
\end{equation} 
and the running coupling depending on the system temperature, 
\begin{equation} 
\frac{1}{g^2(T)} = 2b_0 \ln \frac{T}{\Lambda} + 
\frac{b_1}{b_0} \ln ( 2\ln \frac{T}{\Lambda} ), 
\end{equation} 
where $c$ and $\Lambda$ are free parameters for fitting.
Using the data for $T/T_c=1.5-5.0$ we obtained 
the fitted result, listed in TABLE \ref{tab2}, and in particular, 
the fitted line using the data for $T/T_c=2.0 - 4.0$
is shown as the solid line in Fig. \ref{CStringatfinite}.
It is found that the color-Coulomb string tension
in the thermal phase is described by the magnetic scaling. 

The fitted results in the present lattice simulation  
 seem to depend significantly on the fitting condition. However, 
the fitted value of the coefficient by the magnetic scaling is comparable
 with that reported by the following studies:
$c=0.566(13)$ and $\Lambda/T_c=0.104(9)$
 by the lattice calculation of the spatial Wilson loop in Ref. \cite{SWL6},
$c=0.554(04)$
by the numerical study of the 3-D $SU(3)$ gauge theory in Ref. \cite{SWL8},
and $c=0.482(31)-0.549(16)$ 
from the lattice calculation of the spatial gluon propagator
 in Ref. \cite{SGluon}.
Furthermore, the analysis of the magnetic mass
 in a self-consistent way of high QCD theory leads to 
$c\sim 0.569$ \cite{Nair}.
If one requires more precise data for 
magnetic (spatial) or long-range quantities, 
which are expected to be sensitive to the volume size \cite{SGluon},
 then the computation on the larger lattice is necessary.

In addition, if we assume the electric scaling given by Eq. (\ref{ele})
to describe the thermal color-Coulomb string tension, then  
 the function $1/c g(T)$ is employed.
The results are listed in TABLE \ref{tab2} and the resultant line 
obtained in the fitted range of $T/T_c=2.0-4.0$
is shown as the dashed line in Fig. \ref{CStringatfinite}. 
It seems that the electric scaling also yields
 a good description.
 In the perturbation theory of the leading order, 
the coefficient of the electric scaling is known to be $c=1$ \cite{TQCD},
 and moreover the nonperturbative lattice simulation gives 
the electric scaling with $c > 1$ \cite{SGluon}.
Thus, this procedure may not be as proper a way as the analysis based on 
magnetic scaling.
Nonetheless, this implies that 
in the temperature range of $T/T_c = 1.5 - 5.0$,
the magnitude of the coupling constant is of order 1, 
i.e. there still remains a strong nonperturbative effect.
As a result, the distinction between $g^2(T)$ and $g(T)$  
is not so clear from the present numerical data. 

\begin{table}[h]
\caption{
 The status on the fitting analyses 
of the color-Coulomb string tension at $T \neq 0$.
The second and last rows stand for the fitting range 
in units of $T/T_c$ and the value of the reduced chi-square, 
respectively.
 }
\begin{center}
\begin{tabular}{ccccc}
\hline
\hline
Scaling&Range($T/T_c$) & $c$ & $T_c/\Lambda$ & $\chi^2/ndf$ \\
\hline
Magnetic &2.0 - 5.0 & 0.710(13) & 4.05(20) & 4.45 \\
&2.0 - 4.0 & 0.735(18) & 4.41(29) & 1.47 \\
&1.5 - 4.0 & 0.770(12) & 5.05(20) & 1.99 \\
\cline{2-5}
Electric&2.0 - 5.0 & 0.806(07) & 1.36(3) &  6.30 \\
&2.0 - 4.0 & 0.829(10) & 1.44(4) &  1.25 \\
&1.5 - 4.0 & 0.869(07) & 1.66(3) &  5.52 \\
\hline \label{tab2}
\end{tabular}
\end{center}
\end{table}

\section{Conclusions}

We have investigated the scaling behavior of the color-Coulomb string 
tension in the confinement and deconfinement phases using 
quenched $SU(3)$ lattice gauge simulations. 
The color-Coulomb potential, defined by the Faddeev-Popov operator, 
is an important quantity in discussing the confinement scenario
 in Coulomb gauge, and also from a phenomenological point of 
view.
We have confirmed the scaling behavior of the instantaneous 
color-Coulomb string tension in the confinement and deconfinement 
phases.

In the confinement phase, 
the instantaneous color-Coulomb potential 
behaves as a linearly rising potential at large distances. 
As a consequence, there exists
 the nonvanishing color-Coulomb string tension
 for several coupling constants ($\beta$'s) investigated in this work,
the values of which
are approximately 2 times as large as that of the Wilson loop string
tensions.
 The variation of color-Coulomb string tension on the 
lattice cutoff is also found to be small 
 although we are still far from the 
continuum limit.
These results are qualitatively consistent with those obtained by  
the same analysis in the $SU(2)$ lattice calculations 
in Refs. \cite{Greensite,Greensite2}.
Note that if one employs the gluon propagator itself 
to extract the instantaneous part, then 
the value of the color-Coulomb string tension tends to become smaller 
than that measured by using the Polyakov loop correlator
 as used in the present work \cite{SFGP1}.
However, it is concluded in Ref. \cite{SFGP2} that 
the emergence of a large string tension 
is not ruled out. 

Even in the deconfinement phase, it is observed that 
the color-Coulomb string tension remains finite.
This may be an acceptable result because 
the instantaneous part is constructed in terms of 
 the Faddeev-Popov matrix 
with the derivative operator for the spatial direction; i.e., 
the instantaneous part is not sensitive to the system temperature.
Nevertheless, we should also note that 
the thermal fluctuation still produces a color-screened dynamics 
as has been reported in Ref. \cite{PPLC3}.

The remarkable feature
 that the color-Coulomb string tension does not disappear 
in the deconfinement phase was first shown in the 
$SU(2)$ lattice calculation
\cite{Greensite2} done by Greensite, Olejn\'{i}k and Zwanziger. 
 It is confirmed in the present study   
 that the $SU(3)$ gauge theory has the same feature, 
and we investigated extensively 
the temperature dependence of the color-Coulomb string tension, 
which is found to be in proportion to the temperature.
Note that this issue is supported through
 the discussion of the remnant symmetry in Coulomb gauge 
\cite{Greensite2,Marinari}.

The occurrence of a confining force in the deconfinement phase
was observed in other studies.
The existence of the spatial string tension above $T_c$ is 
well known. In addition,
 as reported in Ref. \cite{SWL7}, 
by the $SU(2)$ lattice simulation in maximally abelian gauge,  
the spatial Wilson loop can almost be reproduced by the 
wrapped monopole loops.
In particular, the 3-D reduction arguments support  
these phenomena.
For the case of the Coulomb gauge QCD, 
the confining linearity in the deconfinement phase 
is caused by the instantaneous part.

In both confinement and deconfinement phases, 
there is no qualitative change in the behavior of the instantaneous part.
Therefore, it is evident that 
the color-Coulomb string tension obtained from that potential 
is not an order parameter for the QGP phase transition.

It is found that the thermal behavior of the color-Coulomb string tension 
is understood by assuming the magnetic scaling, $\sim g^2(T)T$, 
which is actually identified as an infrared regulator or 
a pole mass of the spatial gluon propagator.
If this is a possible interpretation, then 
we can mention the following two points.
Firstly, we conclude that the color-Coulomb string tension in the 
deconfinement phase 
is a kind of the thermal quantity and survives in the high temperature
limit 
as being the same as the case of the spatial Wilson loop.
Secondly, because the magnetic scaling originates in the 
infrared sensitivity of the thermal QCD, in the case of Coulomb gauge, 
the instantaneous part of the gluon propagators reveals 
such infrared behavior.

In the Coulomb gauge confinement scenario  
discussed by Gribov and Zwanziger \cite{Gribov,CCG}, 
the linearity of the instantaneous part for large quark separations
is conjectured to be ascribed to a singularity
 emerging from  the gauge configurations
with a low-lying eigenvalue of the Faddeev-Popov operator.
Hence, it is an important task that 
the distribution of the eigenvalues of the Faddeev-Popov operator
in Coulomb gauge is investigated by the lattice simulation. 
The $SU(2)$ lattice simulation in Coulomb gauge
 performed in Ref. \cite{FP1} proves  
the indication of the enhancement of the low-lying eigenvalues.
The $SU(3)$ lattice study along this line 
is also being undertaken.

In the present work, we focused the calculation of the instantaneous
part only and did not deal with the 
vacuum polarization (retarded) part, being of little significance 
in the view of understanding color confinement in Coulomb gauge. 
However, the vital change concerning the QGP phase transition 
seems to be relevant to the vacuum polarization part,
 the role of which ought to be discussed in a subsequent study. 

In a phenomenological point of view,
 it is interesting that there remains the thermal string tension. 
 If this is regarded as the indication that 
 confining features survive above the critical temperature,
 then this observation may provide some insight into understanding 
 the strongly correlated QGP.  
It has been tried in Ref. \cite{EOSfFMR} 
the description of equation of state in  
the quasiparticle model with the dispersion relation of Gribov type.
However, in other cases,  
it is not obvious how the confining property in the thermal phase
affects physical spectroscopy.
Nevertheless, in Coulomb gauge, it is significant that
 these findings are achieved by classifying 
the time-time gluon propagator into the instantaneous and
 noninstantaneous parts.

\section{Acknowledgments} 

The simulation was performed on an SX-5(NEC) vector-parallel computer 
at the RCNP of Osaka University. 
We appreciate the warm hospitality and support of
 the RCNP administrators.
This work is supported by Grants-in-Aid for Scientific Research from
Monbu-Kagaku-sho (Nos. 13135216 and 17340080).

\end{document}